\documentclass[conference]{IEEEtran}
\IEEEoverridecommandlockouts

\usepackage{cite}
\usepackage{amsmath,amssymb,amsfonts}
\usepackage{algorithmic}
\usepackage{graphicx}
\usepackage{textcomp}
\usepackage{xcolor}
\usepackage{authblk}
\usepackage{textcomp}
\usepackage{xcolor}
\usepackage{booktabs} 
\usepackage{array}
\usepackage{multirow}
\usepackage{booktabs}

\def\BibTeX{{\rm B\kern-.05em{\sc i\kern-.025em b}\kern-.08em
    T\kern-.1667em\lower.7ex\hbox{E}\kern-.125emX}}
\begin{document}

\title{Efficient Long Speech Sequence Modelling for Time-Domain Depression Level Estimation\\
{}
\thanks{}
}

\author[]{\textit{Shuanglin Li}}
\author[]{\textit{Zhijie Xie}}
\author[]{\textit{Syed Mohsen Naqvi}}

\affil[]{Intelligent Sensing and Communications Research Group, Newcastle University, UK}

\maketitle

\begin{abstract}
Depression significantly affects emotions, thoughts, and daily activities. Recent research indicates that speech signals contain vital cues about depression, sparking interest in audio-based deep-learning methods for estimating its severity. However, most methods rely on time-frequency representations of speech which have recently been criticized for their limitations due to the loss of information when performing time-frequency projections, e.g. Fourier transform, and Mel-scale transformation. Furthermore, segmenting real-world speech into brief intervals risks losing critical interconnections between recordings. Additionally, such an approach may not adequately reflect real-world scenarios, as individuals with depression often pause and slow down in their conversations and interactions. Building on these observations, we present an efficient method for depression level estimation using long speech signals in the time domain. The proposed method leverages a state space model coupled with the dual-path structure-based long sequence modelling module and temporal external attention module to reconstruct and enhance the detection of depression-related cues hidden in the raw audio waveforms. Experimental results on the AVEC2013 and AVEC2014 datasets show promising results in capturing consequential long-sequence depression cues and demonstrate outstanding performance over the state-of-the-art.
\end{abstract}


\section{Introduction}
Depression is a mental disorder that significantly disrupts daily life and work accompanied by persistent negative emotions that severely impair both physical and psychological well-being \cite{otte2016major, Nash2024}. Early detection and intervention are crucial to alleviating this suffering \cite{nash2023machine}. However, the diagnostic process is demanding and requires significant effort from experienced psychiatrists, often resulting in delays that prevent many patients from receiving timely treatment \cite{li2024harnessing, li2024acoustic}.

Research has shown that speech signals differ significantly between depressed and non-depressed individuals \cite{he2022deep,france2000acoustical}. Building on these insights, recent studies on depression assessment have utilized time-frequency (T-F) representations from speech and employed neural networks to capture both temporal and spectral patterns \cite{niu2019automatic, dong2021hierarchical}. Recurrent neural networks (RNNs) are effective in modelling temporal dependencies but struggle with gradient vanishing and explosion, and require sequential data processing. Conversely, convolutional neural networks (CNNs) perform well at capturing local features but face limitations in handling long-range dependencies due to their fixed kernel sizes, and their varying pooling operations can introduce biases in the depression estimation task \cite{dong2021hierarchical, niu2019automatic, xie2023position}.

Nevertheless, the efficacy of traditional T-F representations like Fourier and Mel spectrograms in speech-based affective state estimation is increasingly questioned \cite{chen2023speechformer++, niu2023wavdepressionnet}. Emerging studies suggest that directly processing raw speech signals in the time domain offers superior performance in capturing affective information, indicating that raw waveform processing may be more effective for extracting mental disorder-related features. While raw waveform input offers richer information, it poses challenges such as the curse of dimensionality and growing non-stationary with longer speech \cite{niu2023wavdepressionnet, rabiner1993fundamentals}.

Moreover, current methods for assessing depression through speech often segment recordings into short intervals to manage variability in real-world data \cite{cai2021tdca, dong2021hierarchical, fan2022csenet}. However, relying solely on these short segments fails to accurately reflect clinical settings, where longer speech samples are crucial. These extended samples capture the comprehensive speech patterns and emotional cues needed to assess depression accurately.

Recent advancements in long-sequence modelling have been powered by innovations such as DPRNN \cite{luo2020dual}, which uses bidirectional RNNs (Bi-RNN) to process speech in chunks, and by integrating transformers within a dual-path architecture to enhance efficiency \cite{subakan2021attention, chen2020dual}. State Space Models (SSMs) have further streamlined speech processing by reducing computational loads and managing long-range dependencies \cite{gu2021combining, gu2022efficiently}. Mamba\cite{gu2023mamba}, a specialized SSM, processes complex waveforms effectively \cite{gu2023mamba} but faces challenges with quadratic complexity when integrated with traditional attention mechanism \cite{zhang2024mamba}. Recently, the External Attention (EA) \cite{guo2022beyond} mechanism addresses this limitation by using external memory for scalability and simplifying the computation of inter-sequence similarities.

\begin{figure*}
    \centering
    \includegraphics[scale=0.52]{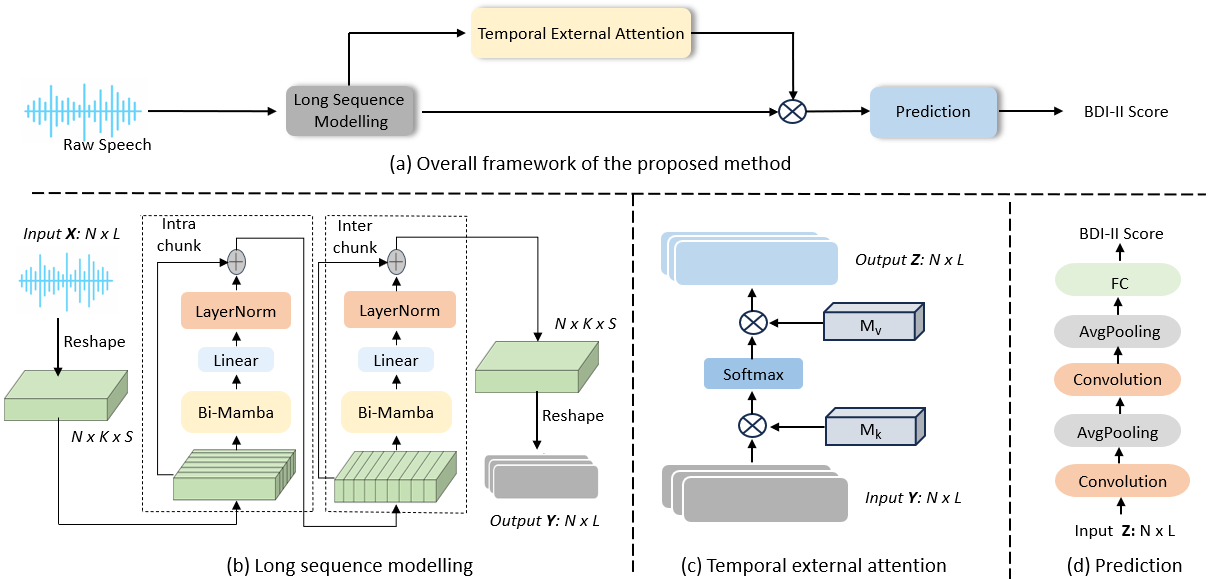}
    \caption{(a) Overall framework of the proposed methods which consists of the long sequence modelling module, the temporal external attention module, and the prediction module. (b) Illustration of the long sequence modelling module. It employs the Bi-Mamba and dual-path architecture to reconstruct the raw audio waves. (c) Illustration of the temporal external attention module. It enhances depression patterns through correlation analysis with additional parameters. (d) Illustration of the prediction module. It predicts the individual depression level scores based on previous output. }
\end{figure*}

Drawing on these insights, we design a network for efficient long-sequence modelling tailored for time-domain depression level estimation, the proposed method incorporates the EA mechanism with bi-directional Mamba (Bi-Mamba) to enhance depression assessment in long speech sequences. The framework comprises a time-domain long-sequence modelling module by substituting the Bi-RNN in DPRNN with Bi-Mamba for improved long sequence processing; a temporal external attention module that uses EA \cite{guo2022beyond} to emphasize depression-related cues; and a prediction module that evaluates depression levels from the reconstructed sequences.

The main contributions of this work can be summarized as follows: 1) We propose an efficient long-sequence modeling network for the depression estimation task. 2) To the best of our knowledge, this is the first study attempt to adopt long speech sequences to reflect the real-world scenarios for depression estimation. 3) The experimental results on the real datasets e.g., AVEC2013 \cite{valstar2013avec} and AVEC2014 \cite{valstar2014avec} demonstrate the effectiveness of the proposed method.

\setlength{\abovedisplayskip}{5pt}
\setlength{\belowdisplayskip}{5pt}
\section{Proposed Method}
In this section, we first review the preliminaries of the state space model which is the basic component of Bi-Mamba. We then provide a detailed explanation of the long-sequence modeling module, the temporal external attention module, and the prediction module.

\subsection{Long Sequence Modelling Module}
The long sequence modelling module has a dual-path structure, integrating SSMs with Mamba. Let $\mathbf{X} \in \mathbb{R}^{N \times L}$ represent the input raw waveform, where $N$ is the feature dimension and $L$ is the sequence length. Following the approach suggested by \cite{luo2020dual}, the raw waveform is transformed into a 3D tensor $\mathbb{R}^{N \times K \times S}$, where $K$ is the chunk length and $S$ is the number of chunks. The output of this module $\mathbf{Y} \in \mathbb{R}^{N \times L}$ maintains the same shape as the input.

\subsubsection{State Space Model}
The SSM maps sequences from $\mathbf{x}(t) \in \mathbb{R}$ to $\mathbf{y}(t) \in \mathbb{R}$ through a hidden state $\mathbf{h} \in \mathbb{R}^{H}$. It utilizes matrices $\mathbf{A} \in \mathbb{R}^{H \times H}$ for state transitions, $\mathbf{B} \in \mathbb{R}^{H \times 1}$ for input projection, and $\mathbf{C} \in \mathbb{R}^{1 \times H}$ for output projection, where \textit{H} denotes the dimension of the hidden state.
\begin{equation}
    \mathbf{h}'(t) = \mathbf{A}\mathbf{h}(t) + \mathbf{B}\mathbf{x}(t), \quad \mathbf{y}(t) = \mathbf{C}\mathbf{h}(t)
\end{equation}

\noindent
For discrete-time signal analysis, the continuous SSM is discretized using matrices $\hat{\mathbf{A}}$, $\hat{\mathbf{B}}$:
\begin{equation}
    \mathbf{h}_t = {\hat{\mathbf{A}}} \mathbf{h}_{t-1} + {\hat{\mathbf{B}}} \mathbf{x}_t, \quad \mathbf{y}_t = \mathbf{C} \mathbf{h}_t
\end{equation}

\noindent
where matrices $\hat{\mathbf{A}}$ and $\hat{\mathbf{B}}$ are derived via a zero-order hold approximation and a learnable parameter $\Delta$ adjusts the balance of current state influence and input at each timestep $\textit{t}$. The Mamba model, an extension of the standard SSM, is designed to be input-selective and dynamically update the parameters $\Delta$, $\hat{\mathbf{A}}$, $\hat{\mathbf{B}}$, and $\mathbf{C}$ based on input $\mathbf{x}_t$ at each time step $t$, thereby enabling efficient handling of dynamic updates and enhancing input context awareness.

\subsubsection{Dual-path Long Sequence Modelling}
Figure 1(b) shows the long sequence modelling module is composed of two consecutive processing blocks: the intra-chunk and the inter-chunk blocks. Each block initially splits the input tensors and then processes them through the Bi-Mamba followed by the linear layer and the layer normalization, respectively. The detail of Bi-Mamba is depicted in Figure 2 which incorporates two parallel convolutions coupled with SSMs to facilitate bidirectional processing. 
\begin{figure}
    \centering
    \includegraphics[scale=0.40]{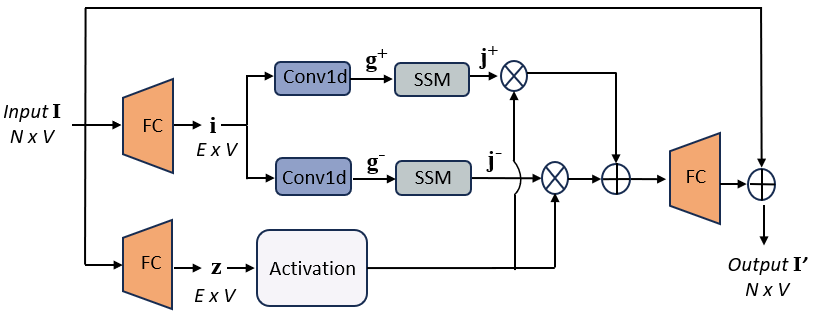}
    \caption{The Bi-Mamba network structure. $\textbf{g}^{\mathbf{+}}$ and $\textbf{j}^{\mathbf{+}}$ stand for the anterior processed sequence. $\textbf{g}^{\mathbf{-}}$ and $\textbf{j}^{\mathbf{-}}$ stand for the posterior processed sequence.}
\end{figure}

\noindent
The original reshaped 3D tensor $\mathbb{R}^{N \times K \times S}$ is initially split along its second and third dimensions into multiple 2D tensors. These tensors are then further categorized into \(S\) sub-intra-tensors of dimensions \(\mathbb{R}^{N \times K}\)  which feed into the intra-chunk block, and \(K\) sub-inter-tensors of dimensions \(\mathbb{R}^{N \times S}\) processed by the inter-chunk block. For clarity of discussion, we denote each input and output of Bi-Mamba as \(\mathbf{I} \in \mathbb{R}^{N \times V}\) and \(\mathbf{I}' \in \mathbb{R}^{N \times V}\), where \(\mathbf{V} \in \{K, S\}\),  respectively.

\noindent
The input sequence \( \mathbf{I} \) is initially processed by a fully connected layer to generate the intermediate representations \(\mathbf{i}\) and \(\mathbf{z}\), both situated in \( \mathbb{R}^{E \times V} \) where \( E=2N \), formulated as:
\begin{equation}
\mathbf{i} = \textit{FC}(\mathbf{I}), \quad \mathbf{z} = \textit{FC}(\mathbf{I})
\end{equation}
\noindent
Besides, to achieve the bidirectional functionality, the intermediate representation  \(\mathbf{i}\) undergoes further processing through two parallel convolution layers for anterior and posterior processing:
\begin{equation}
\textbf{g}^{\mathbf{+/-}} = \sigma(\text{Conv}_{1d}(\mathbf{i}))
\end{equation}
\noindent
where $\sigma$ represents the sigmoid activation function. The term $\textbf{g}^{\mathbf{+}}$ denotes the anterior processed output, while $\textbf{g}^{\mathbf{-}}$ refers to the posterior processed output. The anterior and posterior outputs are then fed into the SSM where the SSM output $\textbf{j}^{\mathbf{+/-}}$ is gated by \(\mathbf{z}\):
\begin{equation}
\textbf{j}^{\mathbf{+/-}} = \sigma(\textbf{z}) \otimes \text{SSM}(\textbf{g}^{\mathbf{+/-}})
\end{equation}
\noindent
where $\otimes$ denotes the Hadamard product operator. Finally, the \(\mathbf{I}'\) is obtained by a linear projection of the average of the anterior and the posterior processed sequences \(\textbf{j}^{\mathbf{+/-}}\), after the posterior sequence $\textbf{j}^{\mathbf{-}}$ is swapped back to the original direction:
\begin{equation}
\mathbf{I}' = \textit{FC}\left(\frac{\textbf{j}^{\mathbf{+}} + \text{swap}(\textbf{j}^{\mathbf{-}})}{2}\right)
\end{equation}
\noindent
The $\mathbf{I}'$ is then transformed by the feature dimension back to that of $\mathbf{I}$ with a linear layer followed by a layer normalization and finally reshaped back to $\mathbf{Y} \in \mathbb{R}^{N \times L}$ which share the same dimension as the input $\mathbf{X}$.

\subsection{Temporal External Attention Module}
The EA unit mitigates a major flaw of self-attention, its focus on long-range dependencies while ignoring inter-instance correlations. It replaces the traditional \textit{Keys} and \textit{Values} with learned external parameters, \(M_k\) and \(M_v\), to capture these correlations. In our temporal external attention module, the input \(\mathbf{Y}\) serves as the Query, similar to conventional self-attention, but utilizes \(M_k\) and \(M_v\) instead of \textit{Keys} and \textit{Values}. As illustrated in Figure 1(c), the external attention maps \(M_k\) and \(M_v\) are obtained using the following equations and are sequentially applied to the input sequence with the softmax function.
\begin{equation}
M_k = \Psi\left(\mathbf{L}_{\text{K}}(\mathbf{Y})\right) \in \mathbb{R}^{N \times L}
\end{equation}
\begin{equation}
M_v = \mathbf{L}_{\text{V}}(\mathbf{Y} \times M_k) \in \mathbb{R}^{N \times L}
\end{equation}
\noindent
where \( \mathbf{L}_{\text{K}}(\cdot) \) and \(\mathbf{L}_{\text{V}}(\cdot)\) are linear layers to replace \textit{Keys} and \textit{Values} in the self-attention mechanism and \( \Psi(\cdot) \) is the ReLU activation function.

\subsection{Prediction Module}
The prediction module processes outputs from preceding layers which take the input \(\mathbf{Z} \in \mathbb{R}^{N \times L}\), to compute a depression score. As illustrated in Figure 1(d), this module includes convolution layers, ReLU activation, average pooling, and fully connected layers, which collectively synthesize information and reduce dimensionality to manage model complexity effectively.

\section{Experiments and Results}

\subsection{Dataset}
We conducted our experiments using the well-known AVEC2013 \cite{valstar2013avec} and AVEC2014 \cite{valstar2014avec} datasets. AVEC2013 contains 150 video clips from 84 subjects, each performing 14 tasks with recording durations ranging from 20 to 50 minutes. AVEC2014, derived from AVEC2013, includes the Northwind and FreeForm tasks, each also consisting of 150 video clips. Both datasets are divided into training, development, and test sets, with 50 samples per set. For this study, these tasks were merged, resulting in 100 videos per set. Both datasets are annotated with Beck Depression Inventory-II (BDI-II) scores \cite{mcpherson2010narrative}, which range from 0 to 63. These scores are used to assess depression severity: 0–13 indicates no depression, 14–19 indicates mild depression, 20–28 indicates moderate depression and 29–63 indicates severe depression.

In our experiments, we use the training sets to develop the model, the development sets for tuning and validation, and the test sets to evaluate the final performance and compare it against existing methods.

\subsection{Evaluation Metrics}
In this work, we use root mean square error (RMSE) and mean absolute error (MAE) as the evaluation metrics to evaluate the performance of the proposed method.

\begin{equation}
    RMSE = \sqrt{\frac{1}{N} \sum_{i=1}^{N} (y_i - \hat{y}_i)^2}
\end{equation}
\begin{equation}
    MAE = \frac{1}{N} \sum_{i=1}^{N} |y_i - \hat{y}_i|
\end{equation}
where \(y_i\) and \(\hat{y}_i\) denote the true and predicted BDI-II scores for the \(i\)-th sample, respectively.

\subsection{Experimental Setup}
Firstly, we segmented the original speech recordings into intervals of 15s, 30s, and 50s. The sampling rate is set as 8kHz, and each segment inherits the original BDI-II score label of the corresponding speech recording. We trained all models using a batch size of 1, with the Adam optimizer \cite{kingma2015adam} and a learning rate of 0.002, for 100 epochs. All experiments were conducted on an NVIDIA L40 GPU. The effectiveness of our proposed method on long sequence modelling is compared against three popular methods for long sequence modelling: vanilla transformer\cite{vaswani2017attention}, vanilla Mamba\cite{gu2023mamba} and DPRNN \cite{luo2020dual}. 

\subsection{Comparison with State-of-the-Art}
We initially compared our proposed methods with networks for speech-based depression estimation, by incrementally extending audio inputs from 12 to 30 seconds at a 6-second step size (12s, 18s, 24s, 30s). Note that we simply adopt the original results for two dataset baseline methods.

\begin{figure}
    \centering
    \includegraphics[scale=0.42]{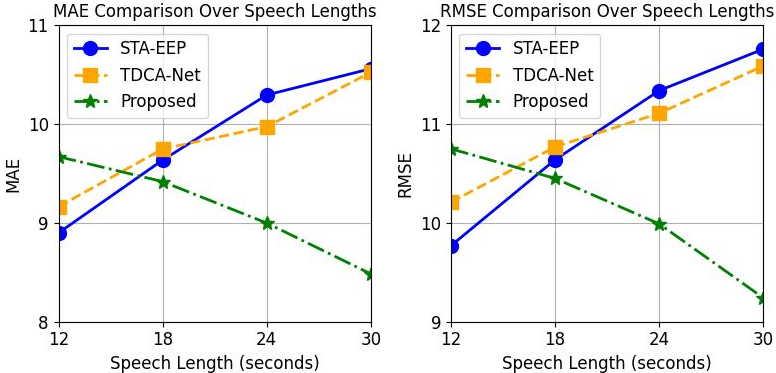}
    \caption{Comparison of network performance based on RMSE and MAE across different speech lengths in the AVEC14 dataset.}
\end{figure}

As depicted in Figure 3 and Table \MakeUppercase{i}, we observed a decline in the performance of existing methods as the speech input length increased. In contrast, our proposed method demonstrated a consistent upward trend, even with longer speech inputs. Notably, the ablation study indicates that the long sequence modelling and the temporal external module within our framework boost the performance, with the temporal external attention outperforming the self-attention, highlighting the effectiveness in handling longer sequence inputs.

\begin{table}[h]
\centering
\caption{Performance comparison on the AVEC 2014 datasets with 30s length input. ``L'' stands for Long Sequence Modelling module, ``TEA'' stands for Temporal External Attention module, and ``SA'' stands for Self-Attention.}
\label{table:performance_comparison}
\begin{tabular}{ccccc}
\toprule
\multirow{2}{*}{\textbf{Methods}} & \multicolumn{2}{c}{\textbf{AVEC2013}} & \multicolumn{2}{c}{\textbf{AVEC2014}} \\
\cmidrule(lr){2-3} \cmidrule(lr){4-5} 
  & RMSE & MAE & RMSE & MAE \\
\midrule
AVEC2013 Audio Baseline \cite{valstar2013avec} & 14.12 & 10.35 & - & - \\
AVEC2014 Audio Baseline \cite{valstar2014avec} & - & - & 12.56 & 10.03 \\
STA-EEP \cite{niu2020multimodal} & 11.76 & 9.82 & 11.79 & 10.73 \\
TDCA-Net \cite{cai2021tdca} & 10.54 & 9.58 & 11.68 & 10.40 \\ \hline
\textit{L} & 9.82 & 8.75 & 9.92 & 8.90 \\
\textit{L} + SA & 9.56 & 8.49 & 9.56 & 8.56 \\
\textit{L + TEA} & \textbf{9.24} & \textbf{8.26} & \textbf{9.20} & \textbf{8.39} \\
\bottomrule
\end{tabular}
\end{table}

\subsection{Comparison with Different Long-sequence Modelling Networks}
To further demonstrate our proposed method on long sequence modelling performance, we replaced the Bi-Mamba in the long sequence modelling module with the vanilla Transformer \cite{vaswani2017attention}, vanilla Mamba \cite{gu2023mamba}, and Bi-RNN \cite{luo2020dual}. Additionally, we compared the proposed method with various popular long-sequence modelling networks, using speech lengths of 15s, 30s, and 50s, to evaluate performance enhancements across extended input speech duration.

\begin{table}[ht]
\centering
\caption{The comparison of experimental performance using different methods on the testing set of AVEC 2013.}
\begin{tabular}{cccc}
\hline
\textbf{Speech Lengths} & \textbf{Methods} & \textbf{RMSE}$\downarrow$ & \textbf{MAE}$\downarrow$ \\
\hline
\multirow{4}{*}{15s} & Transformer \cite{vaswani2017attention} & 11.40 & 10.60 \\
                     & DPRNN \cite{luo2020dual}       & 10.65 & 9.70 \\
                     & Mamba \cite{gu2023mamba}      & 10.15 & 9.15 \\
                     & \textit{Proposed}    & \textbf{9.50} & \textbf{8.75} \\
\hline
\multirow{4}{*}{30s} & Transformer \cite{vaswani2017attention}  & 11.43 & 10.42 \\
                     & DPRNN \cite{luo2020dual}       & 10.40 & 9.71 \\
                     & Mamba \cite{gu2023mamba}       & 10.15 & 8.95 \\
                     & \textit{Proposed}     & \textbf{9.14} & \textbf{8.35} \\
\hline
\multirow{4}{*}{50s} & Transformer \cite{vaswani2017attention} & 11.37 & 10.55 \\
                     & DPRNN \cite{luo2020dual}      & 10.97 & 9.15 \\
                     & Mamba \cite{gu2023mamba}      & 9.60 & 8.72 \\
                     & \textit{Proposed} & \textbf{9.05} & \textbf{8.42} \\
\hline
\end{tabular}
\end{table}

\begin{table}[ht]
\centering
\caption{The comparison of experimental performance using different methods on the testing set of AVEC 2014.}
\begin{tabular}{cccc}
\hline
\textbf{Speech Lengths} & \textbf{Methods} & \textbf{RMSE}$\downarrow$ & \textbf{MAE}$\downarrow$ \\
\hline
\multirow{4}{*}{15s} & Transformer \cite{vaswani2017attention} & 11.51 & 10.74 \\
                     & DPRNN \cite{luo2020dual}      & 10.82 & 9.79 \\
                     & Mamba \cite{gu2023mamba}      & 10.24 & 9.26 \\
                     & \textit{Proposed} & \textbf{9.65} & \textbf{8.83} \\
\hline
\multirow{4}{*}{30s} & Transformer \cite{vaswani2017attention}  & 11.63 & 10.52 \\
                     & DPRNN \cite{luo2020dual}       & 10.72 & 9.26 \\
                     & Mamba \cite{gu2023mamba}        & 10.46 & 8.86 \\
                     & \textit{Proposed} & \textbf{9.20} & \textbf{8.39} \\
\hline
\multirow{4}{*}{50s} & Transformer \cite{vaswani2017attention} & 11.57 & 10.63 \\
                     & DPRNN \cite{luo2020dual}      & 11.17 & 8.92 \\
                     & Mamba \cite{gu2023mamba}      & 9.75 & 8.84 \\
                     & \textit{Proposed} & \textbf{9.14} & \textbf{8.52} \\
\hline
\end{tabular}
\end{table}

As shown in Table \MakeUppercase{ii} \& \MakeUppercase{iii}, the proposed method consistently exhibits superior performance across varying speech durations, surpassing other methods with lower RMSE and MAE scores. This consistent superiority suggests robust architectural and procedural enhancements, potentially involving advanced sequence reconstruction and inter-feature relationship calculation that effectively mitigates error accumulation in longer sequences. These improvements not only emphasize the method's reliability but also enhance its capability to deliver context-rich assessments of depression in real-world scenarios.

\section{Conclusion}
\label{sec:typestyle}
Motivated by studies using speech signals as biomarkers for depression identification and the recent advancement of the long sequence modelling technique, this paper proposed an efficient long sequence modelling network for time-domain depression level estimation using speech. The proposed method provided a long sequence modelling module consisting of a dual-path structure with Bi-Mamba to reconstruct the input raw speech signal, along with additional temporal external attention to enhance the depression cues in the input features by using additional external context. Experiments on the AVEC2013 and AVEC2014 datasets demonstrated that the proposed method achieves promising results even when speech length increases. We planned to extend the work to processing multiple paralinguistic speech-processing tasks in the future. 

\newpage 
\bibliographystyle{ieeetr}
\bibliography{shuanglin}
\end{document}